# Performance Analysis of VoIP Traffic in WiMAX using various Service Classes

Tarik ANOUARI[1]             Abdelkrim HAQIQ[1, 2]

[1] Computer, Networks, Mobility and Modeling laboratory
Department of Mathematics and Computer
FST, Hassan 1st University, Settat, Morocco
[2] e-NGN Research group, Africa and Middle East

## ABSTRACT
Worldwide Interoperability for Microwave Access (WiMAX) is currently one of the hottest technologies in wireless, it's a standard-based on the IEEE 802.16 wireless technology that provides high throughput broadband connections over long distance, which supports Point to Multi-point (PMP) broadband wireless access. In parallel, voice Over Internet Protocol is a promising new technology which provides access to voice communication over internet protocol based network, it becomes an alternative to public switched telephone networks due to its capability of transmission of voice as packets over IP networks. Therefore VoIP is largely intolerant of delay and hence it needs a high priority transmission. In this paper we investigate the performances of the most common VoIP codecs, which are G.711, G.723.1 and G.729 over a WiMAX network using various service classes and NOAH as a transport protocol. To analyze the QoS parameters, the popular network simulator ns-2 was used. Various parameters that determine QoS of real life usage scenarios and traffic flows of applications is analyzed. The objective is to compare different types of service classes with respect to the QoS parameters, such as, throughput, average jitter and average delay.

## 1. INTRODUCTION
Many authors have worked on various QoS parameters using different service classes in WiMAX. A study was conducted on various quality parameters impacting the WiMAX service performance of a WiMAX network. The study suggests that these critical parameters of QoS are required to increase the performance of a WiMAX network. In [4], various simulations using different voice codec schemes and statistical distribution were studied and many performance parameters were analyzed in order to point out the better choice of VoIP codec. Pranita D. Joshi and S. Jangale [7] analyzed various critical QoS parameters like throughput, average jitter and average delay for VOIP using NOAH protocol in NS-2 simulator. Their simulation focuses on the Qos parameters for Best Effort service class only. Similar analysis have been conducted in [3] to examine the QoS deployment over WiMAX network and compare the performance obtained over UGS and ertPS service classes. H. Abid, H. Raja, A. Munir, J. Amjad, A. Mazhar and D. Lee [2] performed a performance analysis when multimedia contents are transferred over WiMax network using Best Effort and ertPS service classes. M. Vikram and N. Gupta [13] analyzed the Qos parameters for WiMAX networks, their performance analysis focuses on UGS service class. The earlier work [7] was limited to analyzing throughput, jitter and delay for Best Effort service flow. Here we have taken more service classes which are rtPS and UGS, more nodes and we have also reproduced the same simulations scenarios to carry out the Qos parameters.

The rest of this paper is organized as follows. Section 2 gives a brief description of WiMAX. Section 3 describes the VoIP technology. The simulation environment and performance parameters are described in Section 4. In Section 5 we present simulation results and analysis. Finally, Section 6 concludes the paper.

## 2. WIMAX OVERVIEW
WiMAX (Worldwide Interoperability for Microwave Access) is a wireless communications standard intended to provide 30 to 40 Mps data rates, providing up to 1 Gbit/s for fixed stations. It is based on IEEE 802.16e-2005 standard [6], which added some improvement to 802.16-2004 standard [5], taking into account mobility. Several new techniques (OFDMA turbo code, FFT, EAP, MIMO ...) are used for a better support for Quality of Service.

It can be used in both point to point (P2P) and the typical WAN type configurations. WiMAX supports different multimedia applications as VoIP, voice conference and online gaming. The IEEE 802.16 technology (WiMAX) is an improved alternative to 3G or wireless LAN networks for providing ease of access, low cost and large coverage area.

### 2.1 Quality of Service in WiMAX Networks
Quality of Service (QoS) [12] is the ability to communicate in good conditions a type of traffic, in terms of availability, throughput, transmission delay, jitter, packet loss, and rate…etc. It has become an important factor to support variety of applications that use network resources. These applications include multimedia services, voice over IP…etc.

The traffic engineering term Quality of service refers to the probability of the telecommunication network meeting a given traffic contract, or the probability of a succeeding packet in the transition between two points in the network.

As the name suggests that it is a measure of how reliable and consistent a network is, the main detractors from respectable QoS are throughput, latency, jitter and percentage of packets lost etc. Resolve these issue and you get a carrier-grade service. The chief goal of a good QoS is to deliver priority including better throughput, controlled jitter and latency (requisite by some real-time and interactive traffic), and enhanced loss characteristics.

## 2.2 WiMAX Network Architecture

The WiMAX network architecture can offers multiple levels of QoS over its classification, queuing, control signaling mechanisms, scheduling, modulation, and routing. It's is a combination of subscribers (SS) and base station (BS). Figure 1 shows the WiMAX network architecture.

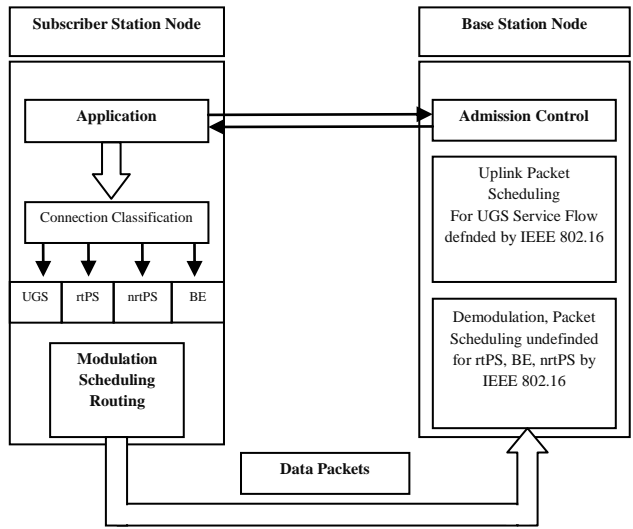

**Fig 1: WiMAX Network Architecture**

## 2.3 QoS Service Classes in WiMAX

WiMAX accords network operators the opportunity to deliver a wealth of services to distinguish their offerings and attract a tiered range of subscribers. It features a diversity of flow types that can be used to improve performance for voice, video, and data. For example, a user having a Voice over IP (VoIP) application needs a real-time data stream unlike another one transferring an FTP file. The IEEE802.16 WiMAX standard offers four categories for the prioritization of traffic named as nsolicited Grant Service (UGS), Real-Time Polling Service (rtPS), Non-Real Time Polling Service (nrtPS), and Best Effort, there is a fifth type QoS service class which is added in 802.16e standard, named as: extended real-time Polling Service (ertPS). IEEE 802.16 has five QoS classes [1].

Table 1 classifies different service classes defined in WiMAX and its description and Qos parameters.

**Table 1. QoS service classes in wimax**

| Service | Description | QoS parameters |
|---|---|---|
| UGS | Support of real-time service flows that generate fiixed-size data packets on a periodic basis, such as VoIP without silence suppression | Maximum sustained rate<br>Maximum latency tolerance<br>Jitter tolerance |
| rtPS | Support of real-time service flows that generate transport variable size data packets on a periodic basis, such as streaming video or audio | Minimum reserved rate<br>Maximum sustained rate<br>Maximum latency tolerance<br>Traffic priority |
| ertPS | Extension of rtPS to support traffic flows such as variable rate VoIP with Voice Activity Detection (VAD) | Minimum reserved rate<br>Maximum sustained rate<br>Maximum latency tolerance<br>Jitter tolerance<br>Traffic priority |
| nrtPS | Support for non-real-time services that require variable size data grants on a regular basis | Minimum reserved rate<br>Maximum sustained rate<br>Traffic priority |
| BE | Support for best-effort traffic | Maximum sustained rate<br>Traffic priority |

## 3. VOIP TECHNOLOGY

### 3.1 VoIP Transport System

VoIP uses a combination of protocols for delivering phone data over networks. Various signaling protocols are used, SIP and H.323 can be regarded as the enabler protocols for voice over IP (VoIP) services [8]. VoIP communications require these signaling systems to setup, control, initiate a session and facilitate real-time data transfer in order to provide clear communications. SIP and H.323 works in conjunction with the Real Time Transport Protocol (RTP) and the User Datagram Protocol (UDP) to transfer the voice stream. Voice data is putted in data packets using the RTP protocol. The RTP packets, enclosed inside the UDP packets, are then transferred to the receiver.

### 3.2 VoIP Codecs

RTP and UDP protocols are the logical choice to carry voice when TCP protocol favors reliability over timeliness. Voice signals are digitally encoded. This means that each voice signal is converted from digital to analog and back. The analog signal is firstly sampled based on a sampling rate of 8 KHz, 8 bits per sample is the most frequently cases. Next, the output is encoded according to many factors: the compression rate and the framing time or the frames length. Finally, one or more of these frames are encapsulated into an RTP/UDP/IP packet for transmission over the network. All these practices are accomplished by one of various audio codecs, each of which vary in the sound quality, the bandwidth required, the computational requirements, encoding algorithm and coding delay [8, 9, 14] :

• **G.711** is the default standard for all vendors, very low processor requirements. This standard digitizes voice into 64 Kbps and does not compress the voice, It performs best in local networks where we have lots of available bandwidth.

• **G.729** is supported by many vendors for compressed voice operating at 8 Kbps. Excellent bandwidth utilization and Error tolerant with quality just below that of G.711.

• **G.723.1** was once the recommended compression standard. It operates at 6.3 and 5.3 Kbps. High compression with high quality audio. Although this standard reduces bandwidth consumption, voice is much poorer than with G.729 and is not very popular for VoIP.

Table 2 shows some features of the most common codecs: G.711, G.723.1 and G.729.

## Table 2. VoIP codecs characteristics

| IUT-T Codec | Algorithm | Codec Delay (ms) | Bit Rate (kbps) | Packets Per Second | IP Packet Size (bytes) |
|---|---|---|---|---|---|
| G.711 | PCM | 0.375 | 64 | 100 | 120 |
| G.729A | ACELP | 35 | 8 | 100 | 50 |
| G.723.1 | CS-ACELP | 97.5 | 5.3 | 33 | 60 |

# 4. SIMULATION ENVIRONNEMENT
## 4.1 Simulation Model
The Network Simulator 2 (NS-2) [10] is a discrete event simulator targeted at networking research, it provides extensive support for simulation of TCP, routing, and multicast protocols over wired and wireless networks. NS is an Object-oriented Tcl (OTcl) script interpreter that has a simulation event scheduler and network component object libraries, it is written in OTcl and in C++, figure 2 illustrates the simulation cycle of NS-2.

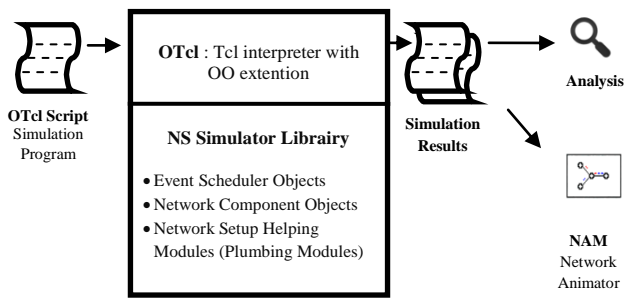

**Fig 2: Simulation Cycle in NS-2**

In this paper, we evaluate the performance of various VoIP codecs using different service classes under NS-2 simulator. Simulation scenario for this study is created such that the mobile nodes are created and associated with the base station they go to at the start. A traffic agent is created and is attached to the source node, a sink node is created and attached to the base station using a wired connection to accept incoming packets.

We have used NS-2 version 2.32 with WiMAX module patch release version 2.6 [11]. The TCL scripts were developed such that the input parameters can be varied. Numbers of mobile nodes in the simulation, type of VOIP codec, and type of service class were passed in as input parameters while running the simulation. The VOIP codecs are varied as G.711, G.723, and G.729. For each of the service class, number of mobile nodes with the VOIP traffic was varied from 2, 4, 6, 8 and 10. Some of the parameters used in simulation are mentioned in table 3. The traffic is started after some time to allow the mobile node to complete the registration because after the simulation starts each node goes through basic registration procedure to get associated with the base station.

The resulted trace files are used for further analysis. Simulation results can be analyzed by running PERL or AWK scripts on the trace file to obtain values of various parameters as throughput, average delay, and average jitter. These analysis results are plotted in graphs to compare following parameters for various service classes.

## 4.2 Simulation Parameters
The simulation parameters are listed in table 3:

**Table 3. Simulation parameters**

| Parameter | Value |
|---|---|
| Simulator | NS-2 (Version 2.32) |
| Channel type | Wireless channel |
| Radio propagation model | Propagation/OFDMA |
| Network interface type | Phy/WirelessPhy/OFDMA |
| MAC type | Mac/802_16/BS |
| Routing protocol | NOAH |
| Antenna model | Antenna/OmniAntenna |
| Link layer type | LL |
| Frame size (msec) | 5 |
| Duplex scheme | TDD |
| Packet Rate | 4 packet/s |
| Modulation Technique | BPSK |
| Simulation time | 100s |

## 4.3 Performance Parameters
### 4.3.1 Throughput
Throughput is the amount of number of packets effectively transferred in a network, in other words throughout is data transfer rate that are delivered to all terminals in a network. It is measured in terms of packets per second or per time slot.

### 4.3.2 Average Delay
Delay or latency represents the time taken by a bit of data to reach from source to destination across the network. The main sources of delay can be categorized into: propagation delay, source processing delay, Queuing delay, transmission delay and destination processing delay. Here we have calculated end to end delay which is a measure of elapsed time taken during modulation of the signal and the time taken by the packets to reach from source to destination.

### 4.3.3 Jitter or Delay variation
Jitter can be observed as the end-to-end delay variation between two consecutive packets. The value of jitter is calculated from the end to end delay. Jitter reveals the variations in latency in the network caused by congestion, route changes, queuing, etc. It determines the performance of network and indicates how much consistence and stable the network is.

# 5. SIMULATION RESULTS AND ANALYSIS
We conducted three sets of simulations, the aim of each one of these sets is to compare the average throughput, average delay and average jitter of Best Effort, rtPS and UGS service classes using VoIP codecs.

The figure 3 shows the graphs of throughput against number of mobile nodes for each codec under various service classes, it is observed that the average throughput increases steadily as the number of nodes increases before reaching six nodes, then it begins to go down.

Among the three graphs, from the sixth node, the throughput of the rtPS class decreases faster than the other service classes and has finally the lowest throughput. Throughputs of BE and UGS traffic are very similar except for the G.711 codec for which the UGS service class performed better than BE.

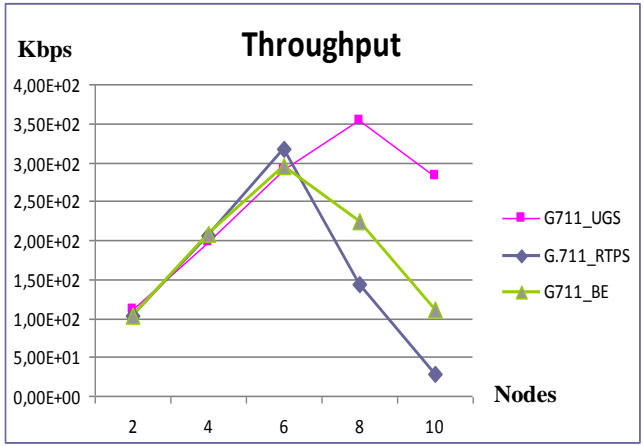

**Fig 3(a): Throughput for G.711 Codec under various service classes**

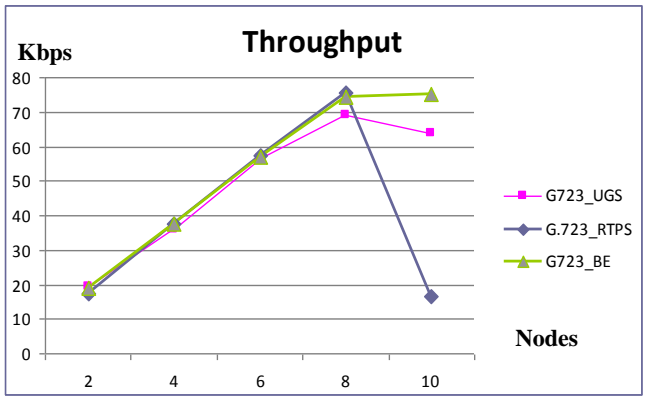

**Fig 3(b): Throughput for G.723 Codec under various service classes**

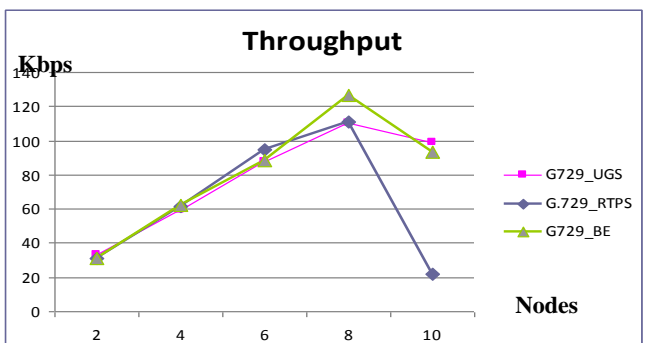

**Fig 3(c): Throughput for G.729 Codec under various service classes**

From the figure 4, BE service class has the highest jitter. Average jitter of all service classes under simulation increases starting from the sixth node, except for the rtPS class which decreases from the eighth node.

In case of the UGS class, the average jitter does not vary as much as the number of nodes increases. In addition to that, the value is very small.

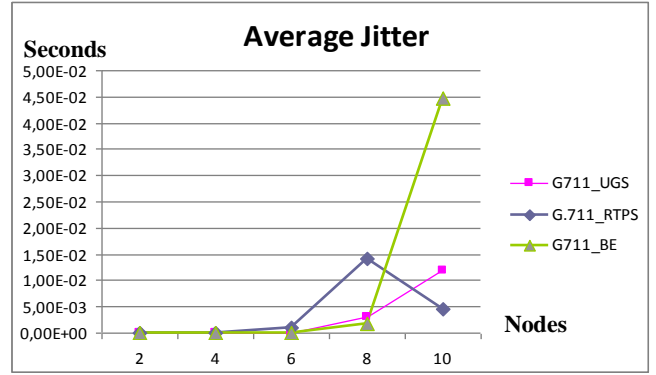

**Fig 4(a): Average Jitter for G.711 Codec under various service classes**

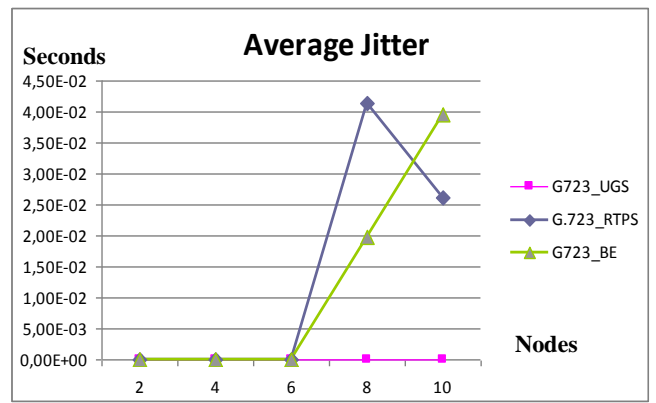

**Fig 4(b): Average Jitter for G.723 Codec under various service classes**

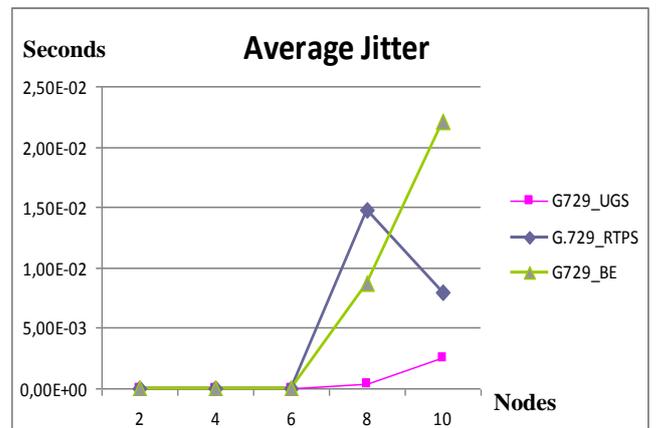

**Fig 4(c): Average Jitter for G.729 Codec under various service classes**

The figure 5 shows the average delay variation of the three service classes. The delay values of BE and rtPS traffic vary similarly with increasing nodes and keep still higher compared with UGS traffic.

From node 2 to 6, average delay is practically naught for the three VoIP codecs, starting from the sixth node, the average delay of rtPS and BE traffic increases sharply. Whereas, the UGS traffic keeps insignificant in comparison to BE and rtPS.

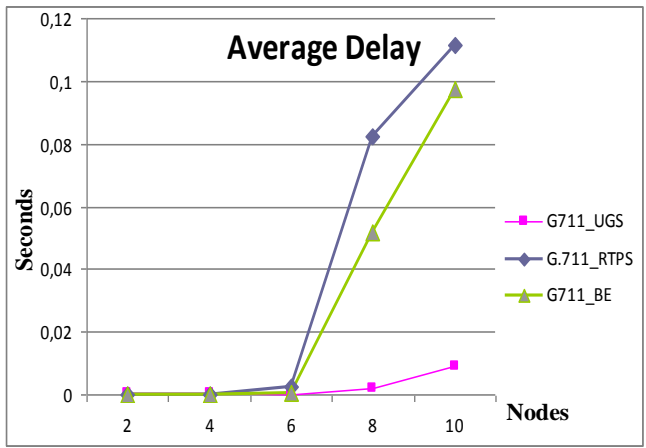

**Fig 5(a): Average Delay for G.711 Codec under various service classes**

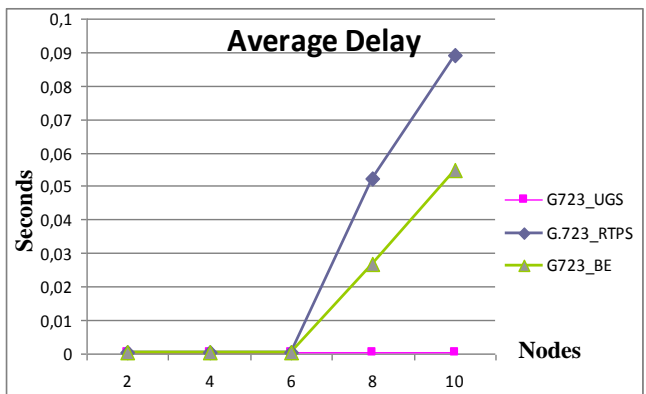

**Fig 5(b): Average Delay for G.723 Codec under various service classes**

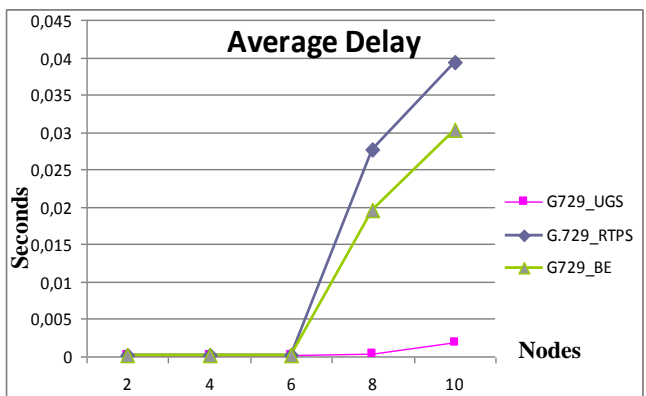

**Fig 5(c): Average Delay for G.729 Codec under various service classes**

## 6. CONCLUSION

In this paper, the BE, rtPS and UGS service classes using different VoIP codecs have been simulated and analysed in terms of throughput, average jitter and average delay.

The rtPS service class comes out to be better than BE service class for average jitter. Otherwise, all the service classes over VoIP codecs under consideration work optimally when it comes to less than six nodes.

In conclusion, it is observed that UGS service class has the best performance parameters serving VoIP. Indeed, UGS service class is designed to handle real-time service flows that generate fixed size packets at regular interval, which is the case for VoIP.

This performance study can be enhanced by treating other type of traffic in instance video on demand VOD under different types of service flows which are supported by fixed WiMAX.

## 7. REFERENCES

[1] Z. Abichar, Y. Peng and J. Morris Chang, "WiMax: The Emergence of Wireless Broadband", IT Professional, Vol. 8, Issue. 4, pp. 44-48, Doi:10.1109/MITP.2006.99, July-Aug. 2006.

[2] H. Abid, H. Raja, A. Munir, J. Amjad, A. Mazhar, D. Lee, "Performance Analysis of WiMAX Best Effort and ertPS Service Classes for Video Transmission", ICCSA, Issue 3, pp. 368-375, 2012.

[3] I. Adhicandra, "Measuring data and VoIP traffic in WiMAX networks", Arxiv Preprint arXiv: 1004.4583, 2010.

[4] S. Alshomrani, S. Qamar, S. Jan, I. Khan and I. A. Shah, "QoS of VoIP over WiMAX Access Networks", International Journal of Computer Science and Telecommunications, Vol. 3, Issue 4, April 2012.

[5] IEEE 802.16-2004, "IEEE Standard for Local and Metropolitan Area Networks Part 16: Air Interface for Fixed Broadband Wireless Access Systems", October, 2004.

[6] IEEE standard 802. 16-2005, "IEEE standard for Local and Metropolitan Area Networks-Part16: Air Interface for Fixed and Mobile Broadband wireless Access systems Amendment 2" February 28, 2006.

[7] D. Joshi, and S. Jangale, "Analysis of VoIP traffic in WiMAX using NS2 simulator" International Journal of Advanced Research in Computer Science and Electronics Engineering, Vol. 1, Issue 2, April 2012.

[8] S. Karapantazis and F.-N. Pavlidou, "VoIP: A comprehensive survey on a promising technology", Computer Networks, Vol. 53, Issue 12, pp. 2050-2090, 2009.

[9] A. Kumar and S. Ganesh Thorenoor, "Analysis of IP Network for different Quality of Service", International Symposium on Computing, Communication, and Control (ISCCC 2009) Proc. of CSIT Vol.1 (2011) © (2011) IACSIT Press, Singapore.

[10] Marc Greis, "Tutorial for Network Simulator NS", http://www.scribd.com/doc/13072517/tutorial-NS-full-byMARC-GREIS.

[11] NIST: IEEE 802.16 module for NS-2 (Feb 2009), http://www.antd.nist.gov/seamlessandsecure/pubtool.shtml#tools.

[12] P. Rengaraju, C.H. Lung, A. Srinivasan, R.H.M. Hafez, "Qos Improvements in Mobile WiMAX Networks", AHU J. of Engineering & Applied Sciences, Vol. 3, Issue 1, pp. 107-118 (2010). © 2009 ALHOSN University.

[13] M. Vikram and N. Gupta, "Performance Analysis of QoS Parameters for Wimax Networks", International Journal of Engineering and Innovative Technology, Vol. 1, Issue 5, May 2012.

[14] T. Zourzouvillys and E. Rescorla, "An introduction to standards-based VoIP: SIP, RTP, and friends", Internet Computing, IEEE, Vol. 14, Issue 2, pp. 69 –73, 2010.